\documentclass[12pt]{article}

\usepackage{graphicx, color}
\usepackage{amsfonts}
\usepackage{amsthm}
\usepackage{amsmath}
\usepackage[top=3cm, bottom=3.25cm, left=2.75cm, right=2.75cm]{geometry}

\newcommand{\beq}{\begin{equation}}
\newcommand{\eeq}[1]{\label{#1}\end{equation}}
\newcommand{\ber}{\begin{eqnarray}}
\newcommand{\eer}[1]{\label{#1}\end{eqnarray}}
\newcommand{\re}[1]{(\ref{#1})}

\newcommand{\half}{\textstyle\frac 1 2}

\newcommand{\bbD}[1]{\mathbb{D}_{#1}}
\newcommand{\bbDB}[1]{\bar{\mathbb{D}}_{#1}}

\newcommand{\bbnab}{\mbox{\hbox{$\nabla$\kern-0.65em\lower.39ex\hbox{${}^{\nabla}$}}}}
\newcommand{\one}{\mathbb{I}}
\def\+{{+\!\!\!+}}
\def\pp{\mbox{\tiny${}_{\stackrel\+ =}$}}

\newcommand{\nn}{\nonumber}
\newcommand{\kah}{K\"ahler~}

\begin{document}
\setcounter{page}{0}
\thispagestyle{empty}
\begin{flushright} \small
UUITP-32/11  \\  YITP-SB-12-04\\ Imperial-TP-2012-CH-1\\
\end{flushright}
\smallskip
\begin{center} \LARGE
{\bf Generalized K\"ahler Geometry in\\$(2,1)$ superspace}
\\[12mm] \normalsize
{\bf  Chris Hull$^{a}$, Ulf~Lindstr\"om$^{b}$, Martin Ro\v cek$^{c}$, \\
Rikard von Unge$^{d}$ and Maxim Zabzine$^{b}$} \\[8mm]
{\small\it
$^a$ The Blackett Laboratory, Imperial College London\\
Prince Consort Road, London SW7 2AZ, U.K.\\
~\\
 $^b$Department of Physics and Astronomy
Uppsala University, \\ Box 516,
SE-751 20 Uppsala, Sweden \\
~\\
$^c$C.N.Yang Institute for Theoretical Physics, Stony Brook University, \\
Stony Brook, NY 11794-3840,USA\\
~\\
$^{d}$Institute for Theoretical Physics, Masaryk University, \\
611 37 Brno, Czech Republic \\~\\}
\end{center}
\vspace{10mm}
\centerline{\bf\large Abstract}
\bigskip
\noindent

Two-dimensional $(2,2)$ supersymmetric nonlinear  sigma models can be described
in $(2,2)$, $(2,1)$ or  $(1,1)$ superspaces. Each description emphasizes different aspects of generalized K\"ahler geometry. We investigate the reduction from $(2,2)$ to $(2,1)$ superspace. This has some interesting nontrivial features arising from the elimination of nondynamical fields. We compare quantization in the different superspace formulations.

\eject
\footnotesize
%\addtocontents{toc}
%\tableofcontents{}
%\end{titlepage}
\normalsize
\eject
\tableofcontents
\section{Introduction}
\setcounter{equation}{0}

Supersymmetric sigma models in two dimensions are useful for investigating complex geometry
because of the constraints that supersymmetry imposes on the target space geometry. The geometry
of models with (2,2) supersymmetry [2] has been explored in
investigations of generalized complex geometry \cite{Hitchin:2004ut},
in particular generalized K\"ahler geometry (GKG) \cite{Gualtieri:2003dx}. A number of interesting mathematical structures have been revealed this way. Examples include: the generalized K\"ahler potential underlying all geometrical objects \cite{Lindstrom:2005zr}; a set of coordinates adapted to the full local description of GKG \cite{Lindstrom:2005zr}; a new gerbe structure related to the global description of GKG \cite{Hull:2008vw}; and a generalization of the Monge-Amp\`ere equation \cite{Hull:2010sn}.

There are different ways of characterizing GKG, and sigma models with various numbers of manifest supersymmetries reflect these. The Lagrangian in $(2,2)$ superspace is the generalized K\"ahler potential, 
whereas the $(1,1)$ superspace Lagrangian is given directly in terms of the metric and $B$-field. Here we shall focus on the $(2,1)$ description where the geometry is governed locally by a one-form. 
This is closely related to the local symplectic description of GKG introduced in \cite{Hull:2008vw}.

Sigma models with  $(2,1)$ supersymmetry have a long but somewhat less well
known history. 
The closely related sigma models with  $(2,0)$ supersymmetry were introduced
and  formulated
in $(1,0)$ superspace in  \cite{Hull:1985jv}. The models with $ (2,1)$
supersymmetry 
were originally studied
in $(1,1)$ superspace  in \cite{Hull:1985zy},  where the conditions on the
target space geometry for the existence
of an additional supersymmetry as well as the conditions for finiteness were
given.
A formulation in $(2,1)$ superspace first appeared
in \cite{Dine:1986by}. An alternative superspace formulation was given in
\cite{Howe:1987qv,Howe:1988cj}.

Our starting point will be the $(2,2)$ model which we  reduce to $(2,1)$.
Once we have the classical GKG model in $(2,1)$ superspace, it is natural to ask about quantum properties.
Here there  are also interesting differences between the $(2,2)$, $(2,1)$ and $(1,1)$ analyses and we investigate these.

Sigma models on Generalized K\"ahler manifolds can of course also be described in $(2,0)$ and $(1,0)$ superspace.
This would presumably lead to further interesting geometrical relations. We will not
pursue this line of research here.
\bigskip

The classical results are in sections \ref{GKG}-\ref{sec5-red}, and the quantum aspects in section \ref{sec6}.  
After a brief introduction to $(2,2)$ superspace in section \ref{sec3-N2}, we describe $(2,1)$ superspace 
and sigma models in section \ref{sec4-N21}.  The reduction from 
$(2,2)$ to $(2,1)$ as well as a discussion of the geometric significance is contained in section \ref{sec5-red}. 
Section \ref{sec6} compares the structure of the one-loop counterterms in the various formulations
and discusses the corresponding differences in the renormalization schemes.

Conventions and background material can be found in \cite{Hull:2008vw, Hull:2010sn}.

\section{Generalized K\"ahler Geometry}
\label{GKG}

The bihermitean geometry discovered in \cite{Gates:1984nk} and reformulated as
Generalized K\"ahler Geometry (GKG)  in \cite{Gualtieri:2003dx} is characterized by the data 
$(M,g, J_{\pm})$ where $M$ is a manifold,  $J_{\pm}$  are two complex structures and $g$ is a metric
 hermitean with respect to both of them.  Moreover, the following integrability conditions are required  
\ber
d^c_{+}\omega_{+}+d^c_{-}\omega_{-}=0~,~ dd_{\pm}^c\omega_{\pm}=0~,
\eer{0}
where $\omega_{\pm}:= gJ_{\pm}$ and $d^c_{\pm}$ is the $i(\bar \partial-\partial)$ operator for the corresponding complex structure. These conditions are equivalent to the 
existence of a closed three-form $H$ where 
\ber
H:= d^c_{+}\omega_{+}=-d^c_{-}\omega_{-}~,
\eer{00}
which is used to construct connections $\Gamma^{(\pm)}$ with torsion $\pm\half g^{-1}H$ that preserve
$J_{\pm}$.
Locally we can always find a potential for the torsion: $H=dB$. We refer to this potential as ``the $B$-field''.
Clearly, it is only defined up to a gauge transformation $\delta B=d\Lambda$, and it is often convenient 
to choose a particular gauge.

There is an alternative local description of GKG, derived in \cite{Hull:2008vw}, which emphasizes the relation to symplectic geometry. We can choose a gauge $B=B_+$ for the $B$-field for which the $(1,1)$
part with respect to $J_{+}$ vanishes, or a gauge
 $B=B_-$ for the $B$-field for which the $(1,1)$ part with respect to
$J_{-}$ vanishes. The GKG geometry can then be
formulated in terms of
\ber
{\cal F}_{+}= \frac{1}{2}(B_+ - g)J_+~, ~~~~{\cal F}_{-}=\frac{1}{2}(B_- + g)J_- ~.
\eer{00000}
 (As explained in \cite{Hull:2008vw}, a more global 
description can be given in terms of gerbe-connections.) Remarkably, ${\cal F}_{\pm}$ 
are closed and non-degenerate, so that locally we may define one-forms $\lambda_{\pm}$:
\ber
{\cal F}_{\pm}=d\lambda_{\pm}~.
\eer{000000}

\section{$(2,2)$ superspace}\label{sec3-N2}

Our point of departure is the full description of GKG in $(2,2)$ superspace. 
As shown in \cite{Lindstrom:2005zr}, away from irregular points the geometry
is completely specified in terms of a generalized K\"ahler potential $K$.
If there are no irregular points, the action for the general $(2,2)$ sigma model
is
\ber\nn
&S=\int d^2\xi ~d\theta ^+d\bar\theta^+d\theta^-d\bar\theta^-~K(\phi,\chi, X_{L}, X_{R})\\[1mm]&=\int d^2\xi~\bbD{+}\bbDB{+}\bbD{-}\bbDB{-}~K(\phi,\chi, X_{L}, X_{R})~.
\eer{11}
The $(2,2)$ algebra is
\ber
\{\bbDB{\pm},\bbD{\pm}\}=2i\partial\pp~,
\eer{12}
and the arguments of $K$ are constrained $(2,2)$ superfields satisfying
\ber\nn
&{\rm (anti)chiral}~: ~~&\bbDB{\pm}\boldsymbol{\phi} =0~,~~~~ \bbD{\pm}\boldsymbol{\bar\phi}=0~,\\[1mm]\nn
&{\rm twisted~(anti)chiral}~:~~ &\bbDB{+}\boldsymbol{\chi} =\bbD{-}\boldsymbol{\chi}=0~,~~~
 \bbD{+}\boldsymbol{\bar\chi} = \bbDB{-}\boldsymbol{\bar\chi}=0~,\\[1mm]\nn
&{\rm left~ semi(anti)chiral}:~~&\bbDB{+}\boldsymbol{X}_l =0~,~~~~\bbD{+}\boldsymbol{\bar{X}}_{\bar{l}}=0~,
\\[1mm]
&{\rm right~ semi(anti)chiral}:~~&\bbDB{-}\boldsymbol{X}_{r} =0~,~~~~\bbD{-} \boldsymbol{\bar{X}}_{\bar r}=0~,~ 
\eer{13}
where we adopt the following notation for superfields 
$\phi = (\boldsymbol{\phi}, \bar{\boldsymbol{\phi}})$,  $\chi = (\boldsymbol{\chi}, \bar
{\boldsymbol{\chi}})$, $X_L = (\boldsymbol{X}_l, \bar{\boldsymbol{X}}_{\bar{l}})$, 
$X_R = (\boldsymbol{X}_r, \bar{\boldsymbol{X}}_{\bar{r}})$. 

\section{$(2,1)$ superspace}\label{sec4-N21}
\subsection{Sigma models}

A $(1,1)$ sigma model $(i=1,...,2n)$ is defined by 
\ber
S=\frac12 \int  d^2\xi ~d^2\theta \,D_+\Phi^i(g_{ij} + B_{ij})D_-\Phi^j~,
\eer{1}
where $\Phi^i$ are unconstrained $(1,1)$ superfields and the $(1,1)$ algebra is given by
\ber
 \{ D_\pm, D_\pm\} = 2i\partial\pp~. 
\eer{extar22}
In \cite{AbouZeid:1997cw}, a $(1,1)$ sigma model $(i=1,...,2n)$ 
with an extra (left) supersymmetry is lifted to $(2,1)$ superspace as $(\alpha = 1,...,n)$
\ber
S=-i\int  d^2\xi~ d\theta ^+d\bar\theta^+d\theta^-(\lambda_\alpha D_- \boldsymbol{\varphi}^\alpha+\bar \lambda_{\bar\alpha} D_-\bar{\boldsymbol{\varphi}}^{\bar\alpha})\equiv
-i\int d^2\xi~  d\theta ^+d\bar\theta^+d\theta^-(\lambda_i D_-\varphi^i)~.
\eer{2}
where $\varphi^i=(\boldsymbol{\varphi}^\alpha, \bar{\boldsymbol{\varphi}}^{\bar\alpha})$ with $\bbDB+ \boldsymbol{\varphi}=0=\bbD+ \bar{\boldsymbol{\varphi}}$. The $(2,1)$ supercovariant derivatives satisfy  
\ber
\{\bbDB{+},\bbD{+}\}=2i\partial_{\+}~,~~~~~~~~~~~ D_-^2= i\partial_=~.
\eer{def2,1-algebra}
The metric $g$ and $B$-field (in a suitable gauge) are given by
\footnote{Ref. \cite{AbouZeid:1997cw} uses a one-form potential  $k$ related
to our $\lambda$ through $ k_\alpha =i\lambda _\alpha$.}
\ber
g_{\alpha\bar\beta}&=&i(\partial_{\alpha}\bar \lambda_{\bar \beta}-\partial_{\bar \beta}\lambda_{\alpha})~,
\nonumber\\[1mm]
B^{(2,0)}_{\alpha\beta}&=&i(\partial_{\alpha}\lambda_{\beta}-\partial_{\beta}\lambda_{\alpha})~,~
\hbox{with}\nonumber\\[1mm]
B&=&B^{(2,0)}+B^{(0,2)}~.
\eer{3}
Alternatively, it is always possible to choose a gauge where
\ber
&~B^{(1,1)}_{\alpha\bar\beta}=i(\partial_{\alpha}\bar \lambda_{\bar \beta}+\partial_{\bar \beta}\lambda_{\alpha})~.
\eer{32}
The target space geometry of a $(2,1)$  supersymmetric sigma model
 corresponds to a complex manifold $(M,J)$ with an hermitian metric $g$ such that 
\ber
 d d^c \omega =0~,
\eer{integrabi}
where $\omega = gJ$. In the mathematics literature such geometry  is called strong KT  (strong \kah with torsion).  
Here $H$ is defined by $H= d^c \omega$.  Locally the geometry can be encoded in terms of the one-form
potential $\lambda$ which appears in the $(2,1)$ action (\ref{2}). The geometry is
invariant under the following symmetries:
\ber
    \lambda_\alpha(\boldsymbol{\varphi}, \bar{\boldsymbol{\varphi}}) ~~~\rightarrow~~~\lambda_\alpha (\boldsymbol{\varphi}, \bar{\boldsymbol{\varphi}})+ \partial_\alpha f(\boldsymbol{\varphi}, \bar{\boldsymbol{\varphi}})
     + l_\alpha (\boldsymbol{\varphi})~.
\eer{syemmetriesoflambda}
These transformations change the lagrangian by terms which vanish when integrated over the full superspace.
If $H$ represents an element of $H^3(M, \mathbb{Z})$ then the transformations of $\lambda$ are related to a gerbe
with connection and the problem  can be analyzed very much along the lines described in \cite{Hull:2008vw}.

The $(2,1)$ superspace field equations follow from varying the action (\ref{2}) with respect to $\varphi^i$;
the chirality constraints imply that we may write the variation as 
\ber
&&\delta\boldsymbol{\varphi}=\bbDB+\delta\boldsymbol{\psi}_-~~~,~~~~\delta\bar{\boldsymbol{\varphi}}
=-\bbD+\delta\bar{\boldsymbol{\psi}}_-
\eer{delvarphi}
for arbitrary unconstrained $(2,1)$ superfield variations $\psi= (\boldsymbol{\psi},\bar{\boldsymbol{\psi}})$.
After integration by parts, the variation of the 
action (\ref{2}) may be written as:
\ber
\delta S=-\int d^2\xi~ d\theta ^+d\bar\theta^+d\theta^-~g_{\alpha\bar\beta}\left(\delta\boldsymbol{\psi}^\alpha_-
\bar\bbnab^{(-)}_+ D_-\bar{\boldsymbol{\varphi}}^{\bar\beta}- \delta\bar{\boldsymbol{\psi}}^{\bar\beta}_-\bbnab^{(-)}_+D_- \boldsymbol{\varphi}^{\alpha}\right)~,
\eer{var2}
where $\bbnab^{(-)}_+$ is the pullback to $(2,1)$ superspace of the connection with torsion:
\ber
\nabla^{(-)}=\nabla^{(0)}-\half g^{-1}H~,
\eer{bbnab}
 where $\nabla^{(0)}$ is the Levi-Civita connection.

\subsection {Vector-fields and couplings}
\label{vcouplings}

Assume that our matter action (\ref{2}) is symmetric under
\ber
\boldsymbol{\varphi}^1 \rightarrow e^{i\gamma}\boldsymbol{\varphi}^1~,~~~~~~~~
\bar{\boldsymbol{\varphi}}^{\bar{1}} \rightarrow e^{-i\gamma}\bar{\boldsymbol{\varphi}}^{\bar{1}}
\eer{phisym}
with $\gamma$ real. Geometrically it corresponds to having an isometry which preserves
the strong KT geometry.
We  gauge this symmetry by turning the parameter $\gamma$ into a chiral superfield and
introducing a $(2,1)$ vector multiplet $(v,A_-)$ which transforms  according to
\ber\nn
&&(e^v)'=e^{i\bar \gamma}(e^v)e^{-i\gamma}~,\\[1mm]
&&A'_-=e^{i\bar \gamma}(A_--iD_-)e^{-i\gamma}~.
\eer{vtf}
Coupling to matter is achieved by introducing gauge covariantly chiral fields
$\boldsymbol{\varphi}^1,\tilde{\boldsymbol{\varphi}}^{\bar{1}}=\bar{\boldsymbol{\varphi}}^{\bar{1}}e^v$
which satisfy chirality constraints with respect to the gauge covariant derivatives
${\cal{D}}_+ = e^{-v}\mathbb{D}_+e^v$, $\bar{\cal{D}}_+ = \bar{\mathbb{D}}_+$, i.e.
$\bar{\cal D}_+ \boldsymbol{\varphi}^1 = {\cal D}_+\tilde{\boldsymbol{\varphi}}^{\bar{1}} = 0$.
The gauged action becomes
\ber
S=-i\int d^2\xi~d\theta ^+d\bar\theta^+d\theta^-
\left[\lambda_1(\boldsymbol{\varphi}^1,\tilde{\boldsymbol{\varphi}}^{\bar{1}}, É... ) 
\nabla_-\boldsymbol{\varphi}^1+
\lambda_{\bar{1}}(\boldsymbol{\varphi}^1,\tilde{\boldsymbol{\varphi}}^{\bar{1}}, É... ) 
\nabla_-\tilde{\boldsymbol{\varphi}}^{\bar{1}} + É... \right]~,
\eer{gac}
where
\ber
\nabla_- \boldsymbol{\varphi}^1 =D_- \boldsymbol{\varphi}^1+iA_- \boldsymbol{\varphi}^1
\eer{covder}
and dots stand for the contribution of other fields (spectators) which
remain unchanged compared to the action (\ref{2}).

\subsection{Additional susy in $(2,1)$}
We may start from the $(2,1)$ sigma model and ask under which circumstances it has
$(2,2)$ supersymmetry, {\it i.e.}, what is the condition for an additional right supersymmetry.
We thus consider the action \re{2} and make an ansatz for an additional supersymmetry
using a real superfield parameter $\epsilon$:
\ber
\delta \boldsymbol{\varphi}^\alpha=\bbDB{+}(\epsilon J^{\alpha}_{-i}D_-\varphi^i)
=\bar\bbnab^{(-)}_{+}(\epsilon  J^{\alpha}_{-i}D_-\varphi^i)~,~~~i:=(\alpha, \bar\alpha)~,
\eer{41}
where $\bar\bbnab^{(-)}_{+}$ is the pullback of the connection with the torsion 
 given by $-\frac12g^{-1}H$, as discussed in section \re{GKG}.
The superfields $\{\varphi^i\}=\{\boldsymbol{\varphi}^\alpha,\bar{\boldsymbol{\varphi}}^{\bar\alpha}\}$ are 
   coordinates for
 the complex structure $J_{+}$,
and are chiral (resp.~antichiral): $\bbDB{+}\boldsymbol{\varphi}^\alpha=\bbD{+}\bar{\boldsymbol{\varphi}}^{\bar\alpha}=0$. 
The second equality in \re{41} follows because $\boldsymbol{\varphi}^\beta$ is chiral ($\bbDB{+}\boldsymbol{\varphi}^\beta=0$)
and $\Gamma^{(+)}$ preserves $J_{+}$ holomorphic indices ($\Gamma^{(+)\alpha}_{j\bar\beta}=0$):
\ber
\bar\bbnab^{(-)}_{+}\boldsymbol{\psi}^\alpha-\bbDB{+}\boldsymbol{\psi}^\alpha
\equiv\Gamma^{(-)\alpha}_{ij}\bbDB{+}\varphi^i\psi^j
=\Gamma^{(-)\alpha}_{\bar\beta j}\bbDB{+}\bar{\boldsymbol{\varphi}}^{\bar\beta}\psi^j
=\Gamma^{(+)\alpha}_{j\bar\beta}\bbDB{+}\bar{\boldsymbol{\varphi}}^{\bar\beta}\psi^j=0~.
\eer{41a}
Complex conjugation gives:
\ber
\delta \bar{\boldsymbol{\varphi}} ^{\bar\alpha}=-\bbD{+}(\epsilon J^{\bar\alpha}_{-i}D_-\phi^i)~.
\eer{41c}

The hermitean metric and $B$-field are given in terms of the vector potentials $\lambda_\alpha$ as in \re{3}. Matching with the $(1,1)$ reduction of the transformations (\ref{41}),(\ref{41c}) implies that the transformation parameter $\epsilon$ obeys
\ber\nn
&(\bbDB{+} -\bbD{+})\epsilon =0, ~~\partial_\+ \epsilon=0\\[1mm]
&(\bbDB{+} +\bbD{+})\epsilon :=2i\epsilon^-~.
\eer{42}

We first consider invariance of the action: we vary (\ref{2}) using (\ref{var2}) and (\ref{41}):
\ber
\delta S=-\int d^2\xi~ d\theta ^+d\bar\theta^+d\theta^-~g_{\alpha\bar\beta}\left((\epsilon J^{\alpha}_{-i}D_- \varphi^i)
\bar\bbnab^{(-)}_+ D_-\bar{\boldsymbol{\varphi}}^{\bar\beta}+ c.c.\right)~.
\eer{varJm}
Because the holomorphic superfields $\boldsymbol{\varphi}$ are chiral and the metric is 
hermitean with respect to $J_{+}$,
the variation of the action can be rewritten in terms of $\omega_{-} =g J_{-}$
\ber
\delta S=-\int d^2\xi~ d\theta ^+d\bar\theta^+d\theta^-\left(\epsilon\,\omega_{-ij}D_-\varphi^j
\bar\bbnab^{(-)}_+ D_-\varphi^{i}-\epsilon\,\omega_{-ij}D_-\varphi^j
\bbnab^{(-)}_+ D_-\varphi^{i}\right)~.
\eer{varomm}
This cannot cancel unless the symmetric part of $\omega_{-}$ vanishes and hence the metric is hermitean
with respect to $J_{-}$; then we have
\ber
\delta S=-\int d\theta ^+d\bar\theta^+d\theta^-~\frac12\left(-\epsilon\,\omega_{-ij}\bar\bbnab^{(-)}_+(D_-\varphi^j
 D_-\varphi^{i})+\epsilon\,\omega_{-ij}\bbnab^{(-)}_+(D_-\varphi^j
 D_-\varphi^{i})\right)~.
\eer{varomma}
Integrating by parts, we find that this vanishes when \re{42} is satisfied and when the connection
$\nabla^{(-)}$ preserves $\omega_{-}$, and hence
\ber
\nabla^{(-)}_kJ_{-j}^{i}=0~.
\eer{43}
Note that this allows us to rewrite the transformations \re{41} as
\ber
\delta \boldsymbol{\varphi}^\alpha
=J^{\alpha}_{-i}\bar\bbnab^{(-)}_{+}(\epsilon  D_-\varphi^i)~,
\eer{41111}
which makes it clear that the $\theta$ independent component of $\epsilon$ 
generates central charge transformations proportional to the field equations
\ber\nn
\bar\bbnab^{(-)}_+D_-\bar{\boldsymbol{\varphi}}^{\bar\alpha }=0~.
\eer{422}
These transformations are thus of interest only off-shell. The additional right supersymmetry 
has parameter $\epsilon^-$.
 
Checking closure of the algebra generated by the transformations \re{41}, 
we find that $J_{-j}^{i}$ is an additional complex structure, 
\ber
J_{-}^2=-\one~,~~~~{\cal N}(J_{-})=0~,
\eer{4223}
where ${\cal N}$ is the Nijenhuis tensor.
In addition, the commutator of the new right supersymmetry \re{41} with the existing left
is proportional to field equations times $-iJ_{-\beta}^{\bar \alpha}$ and $iJ_{-\bar\beta}^{\alpha}$, 
which is just the commutator of the left and right complex structures (recall the $J_{+}$ 
has the canonical form $diag(i,-i)$).

\section{From $(2,2)$ to $(2,1)$ superspace}\label{sec5-red}

 In this section we discuss the reduction from $(2,2)$ to $(2,1)$ models. 
Here  we adopt the following short-hand notations for the 
derivatives of $K$:
\ber
K_C &=& \partial_\phi K = (K_c, K_{\bar{c}}) =( \partial_{\boldsymbol{\phi}} K, \partial_{\bar
{\boldsymbol{\phi}}} K) , \nonumber\\
K_T &=& 
\partial_\chi K = (K_t, K_{\bar{t}})= (\partial_{\boldsymbol{\chi}} K, \partial_{\bar
{\boldsymbol{\chi}}} K), \nonumber\\
K_L &=& \partial_{X_L} K = (K_l, K_{\bar{l}}) = ( \partial_{\boldsymbol{X}_l} K, \partial_{\bar
{\boldsymbol{X}}_{\bar{l}}}K),\nonumber\\
K_R &=& \partial_{X_R} K = (K_r, K_{\bar{r}}) = ( \partial_{\boldsymbol{X}_r} K, \partial_{\bar
{\boldsymbol{X}}_{\bar{r}}}K),
\eer{}
where we suppress all coordinates indices.  Analogously we define the matrices of
double derivatives of $K$, e.g. $K_{l\bar{r}}$ is our notation for the matrix of 
second derivatives
$\partial_{\boldsymbol{X}_l} \partial_{\bar{\boldsymbol{X}}_{\bar{r}}} K$ etc. 
We use matrix and vector notation and suppress all indices.  For further explanations of this notation
 the reader may consult \cite{Hull:2010sn}.

\setcounter{equation}{0}
\subsection{Sigma Models}
Let us consider the reduction to $(2,1)$ superspace of a sigma model in $(2,2)$ superspace with action:
\ber
S = \int d^2\xi~d\theta^+d\bar{\theta}^+d\theta^-d\bar{\theta}^- 
K(\phi,\chi , X_L, X_R)~.
\eer{r1}
Of the $(2,2)$ superspace derivatives, we keep $\mathbb{D}_+,\bar{ \mathbb{D}}_+$ and write

\ber\nonumber
\mathbb{D}_-&=\frac{1}{\sqrt{2}} \left ( D_--iQ_- \right )~,\\[1mm]
\bar{\mathbb{D}}_-&=\frac{1}{\sqrt{2}} \left ( D_-+iQ_- \right )~,
\eer{5}
 where $Q_-$ is defined such that it anticommutes with $D_-$ and moreover  these expressions 
  are compatible  with the definitions (\ref{12}) and (\ref{extar22}). 
We further separate $\theta^-$ into its real and imaginary parts, and reduce to $(2,1)$ by dropping
the dependence on the imaginary part (denoted by a vertical bar). We find
\ber\nn
&&S = i\int d^2\xi~\mathbb{D}_+\bar{\mathbb{D}}_+D_-Q_-K(\phi,\chi, X_L,  X_R)|\\[1mm]
&&=i\int d^2\xi~\mathbb{D}_+\bar{\mathbb{D}}_+D_-
\Big(K_L\psi_{-L} + K_RJD_- x_R + K_CJD_- z - K_TJD_- w \Big) ~,
\eer{3333}
where $J$ is the canonical complex structure $diag(i,-i)$ and where $x_L,\psi_{-L}, x_R,z, w$ are $(2,1)$ superfields  defined according to
\footnote{We can further reduce to $(1,1)$ superspace by removing the
dependence on the imaginary part of $\theta^+$.
We then  find spinor superfields $\psi_+$ together with the reduction
of $\psi_-$; these are auxiliary as $(1,1)$ superfields \cite{Buscher:1987uw}.}
\ber
x_L:= X_{L}|~~,~~~\psi_{-L} =Q_-X_L|~~,~~~x_R:= X_R|~~,~~z:=\phi|~~,~~~w:=\chi|~.
\eer{7}
The $(2,1)$ superfields
$x_L = (\boldsymbol{x}_l,\bar{\boldsymbol{x}}_{\bar l})$ , $\psi_{-L} = (\boldsymbol{\psi}_{-l},\bar{\boldsymbol{\psi}}_{-\bar l})$, $z=(\boldsymbol{z}, \bar{\boldsymbol{z}})$ and $w=(\boldsymbol{w}, \bar{\boldsymbol{w}})$ 
are (anti)chiral 
 $(2,1)$ superfields, respectively:
\ber\nonumber
\bar{\mathbb{D}}_+ \boldsymbol{x}_l=\!\!&0&\!\!={\mathbb{D}}_+\bar{\boldsymbol{x}}_{\bar l}~,{}\\[1mm]\nn
\bar{\mathbb{D}}_+ \boldsymbol{\psi}_{-l}=\!\!&0&\!\!={\mathbb{D}}_+\bar{\boldsymbol{\psi}}_{-\bar l}~,{}\\[1mm]\nn
\bar{\mathbb{D}}_+ \boldsymbol{z}=\!\!&0&\!\!={\mathbb{D}}_+\bar{\boldsymbol{z}}~,{}\\[1mm]
\bar{\mathbb{D}}_+ \boldsymbol{w}=\!\!&0&\!\!={\mathbb{D}}_+\bar {\boldsymbol{w}} {}~,
\eer{8}
and $x_R =(\boldsymbol{x}_r, \bar{\boldsymbol{x}}_{\bar r})$ 
 is an unconstrained $(2,1)$ superfield.  
 In $(2,1)$ superspace $K$ is a function of these $(2,1)$ superfields. 
To find the usual $(2,1)$ superspace action \re{2}, we impose the $\psi_-$-field equations;
\ber
\bar{\mathbb{D}}_+K_\ell={\mathbb{D}}_+K_{\bar\ell}=0~,
\eer{7b}
which we can solve for $x_R$ by introducing chiral superfields $y_l = (\boldsymbol{y}_l, \bar{\boldsymbol{y}}_{\bar l})$:
\ber
K_l= \boldsymbol{y}_l~~,~~~K_{\bar l}= \bar{\boldsymbol{y}}_{\bar l}
\eer{7c}
to find $x_R(x_L, y_L, z, w)$.
As $\boldsymbol{\psi}_{-l}$ is chiral, \re{7b} implies that the $\boldsymbol{\psi}_{-l}$ term does not enter
in the final $(2,1)$ superspace Lagrangian. We are left with
\ber\nn
i\int d^2\xi~\mathbb{D}_+\bar{\mathbb{D}}_+D_-
\Big( K_RJD_-x_R(x_L,y_L, z, w) + K_CJD_-z - K_TJD_-w\Big)~.\\
\eer{ni}
We also need to check that the field equations of $x_R$ impose no new constraints -- they yield
the consistent equation $\psi_{-L}=D_-x_L$.

Introducing the notation
\ber
\boldsymbol{\varphi}^{\alpha} = \left(\!\begin{array}{c}
\boldsymbol{x}_l\\ \boldsymbol{y}_l \\ \boldsymbol{z}\\ \boldsymbol{w}\end{array}\!\right)~,
\eer{r5}
the action may be written as in \re{2} with
\ber
\lambda_\alpha = -\left(\!\begin{array}{c}
-K_RJK_{LR}^{-1}K_{Ll} \\
K_RJK_{lR}^{-1} \\
iK_c-K_RJK_{LR}^{-1}K_{Lc} \\
-iK_t - K_RJK_{LR}^{-1}K_{Lt}
\end{array}\!\right)~.
\eer{r7}
Considering the one-form $\lambda:=\lambda_\alpha d\boldsymbol{\varphi}^\alpha+ \bar \lambda_{\bar \alpha}d\bar{\boldsymbol{\varphi}}^{\bar\alpha}$  and comparing to the expressions in 
\cite{Hull:2008vw} shows that we have recovered the one-form $\lambda^{(+)}$ from \re{000000}, which in $(2,2)$ coordinates 
$X_R,X_L,\phi,\chi$ reads:
\ber
\lambda^{(+)}=-K_RJ dX_R - K_CJ d\phi  + K_T J d\chi~.
\eer{recov}

Because all $(2,1)$ fields are chiral it is straightforward to check that reducing
to $(1,1)$ superspace produces an action
\ber
-\int D_+D_- \; \left[ D_+\varphi^i (J_+)_i^{\;\;k}(d\lambda^{(+)})_{kl} D_-\varphi^l \right]
\eer{godown}
which, using \re{00000} and \re{000000}, is clearly the sigma model \re{1} written in coordinates adapted to $J_+$.

\subsection{Vector multiplets}

In $(2,2)$ superspace there are different vector multiplets that are used to gauge various types of 
isometries \cite{Hull:1991uw}, \cite{Lindstrom:2007vc}, \cite{Lindstrom:2008hx}. 
The K\"ahler vector (or twisted K\"ahler) multiplets that gauge isometries 
in the chiral (or twisted chiral) sector reduce straightforwardly to the $(2,1)$ form described in section \re{vcouplings}. The basic Yang-Mills multiplet in $(2,2)$ supersymmetry consists of a real
unconstrained superfield $V$ transforming as
\ber
e^V \rightarrow e^{i\bar\Lambda} e^V e^{-i\Lambda}~,
\eer{kul1}
where $\Lambda$ is a $(2,2)$ chiral superfield.

Reducing to $(2,1)$ superspace we define the components
\ber\nn
e^V| &=& e^v~, \\[1mm]
e^{-V}Q_-e^V| &=& - 2 A_- -ie^{-v}D_-e^v ~,
\eer{kul2}
with gauge transformations
\ber\nn
e^v &\rightarrow & e^{i\bar\lambda} e^v e^{-i\lambda}~,\\[1mm]\nn
A_- &\rightarrow& e^{i\lambda}\left( A_- -  iD_-\right)e^{-i\lambda}~,\\[1mm]
\bar A_- &\rightarrow& e^{i\bar\lambda}
       \left( \bar A_- - iD_-\right)e^{-i\bar\lambda}~,
\eer{kul3}
where $\lambda$ is a $(2,1)$ chiral superfield and there is a reality constraint
\ber
\bar{A}_- = e^v \left(A_- - iD_-\right)e^{-v}~.
\eer{kul4}
The covariant derivative is $\nabla_-\phi = D_-\phi + iA_- \cdot\xi$ where
$\xi$ is the Killing vector.

The twisted K\"ahler multiplet transforms with twisted chiral parameters  $\tilde{\Lambda}$
\ber
e^{\tilde{V}} \rightarrow e^{i\bar{\tilde{\Lambda}}}e^{\tilde{V}}
e^{-i\tilde{V}}~.
\eer{kul5}
The $(2,1)$ components are defined as
\ber\nn
e^{\tilde{V}}| &=& e^{\tilde{v}}~,\\[1mm]
e^{-\tilde{V}}Q_-e^{\tilde{V}}| &=& ie^{-\tilde{v}}D_- e^{\tilde{v}} + 2\tilde{A}_-~,
\eer{kul6}
where $e^{\tilde{v}}$ and $\tilde{A}_-$ transforms exactly as $e^v$ and
$A_-$ and they satisfy the same reality constraint. That is, both $V$ and $\tilde{V}$
reduce to the same multiplet in $(2,1)$ superspace.

In addition there are the Large Vector Multiplet gauges isometries that act on both chiral and twisted chiral coordinates and the semichiral vectormultiplet that gauges isometries among the semichiral coordinates. These multiplets introduce novel features when reduced and will be treated elsewhere.

\subsection{Comment on generalized K\"ahler geometry and superspace}

The superfields of $(2,1)$ superspace are necessarily complex: they are chiral and 
their complex conjugates are antichiral.
Geometrically this means that they are holomorphic (resp. antiholomorphic)
coordinates that put $J_{+}$ into
its canonical form. All the holomorphic coordinates 
are on an equal footing, but we have singled out one
of the two complex structures $J_{+}$ for preferential 
treatment. Similarly, in $(1,2)$ superspace, $J_{-}$
is diagonalized. 

In contrast, in $(2,2)$ superspace, we must choose a {\em polarization} for the semichiral superfields -- the
coordinates along the symplectic leaves on which $[J_{+},J_{-}]$ is invertible. Along these
leaves, we choose half of the $J_{+}$-holomorphic coordinates, and half of the 
$J_{-}$-holomorphic coordinates to write the generalized K\"ahler potential that is the $(2,2)$ superspace
Lagrange density. Different choices of polarization give rise to different generalized K\"ahler potentials;
of course, they all give rise to the same $(2,1)$ Lagrange density up to holomorphic coordinate 
reparameterizations. The $(2,2)$ superspace description, while requiring a choice of polarization, 
treats the two complex structures on equal footing.

\section{Superspace Counterterms and Renormalization}\label{sec6}
\setcounter{equation}{0}
\subsection{Quantization in $(1,1)$ Superspace}
The $(2,2)$ sigma model can be formulated in $(1,1)$, $(2,1)$ or $(2,2)$ superspaces and in each case can be quantized using the corresponding superspace Feynman rules to obtain superspace counterterms. These are of interest due to their relation to the field equations governing string backgrounds. The one-loop counterterm in $(1,1)$ superspace is given in terms of the Ricci curvature with torsion, and comparing this with the counterterms in $(2,1)$ and $(2,2)$
 superspace gives interesting expressions for the Ricci curvature in terms of potentials. The aim of this section is to explore and exploit these relations.

The $(1,1)$ superspace action is 
\ber
S=\int d^2\xi ~d\theta ^+ d\theta^-  E_{ij} D_+\Phi^i D_-\Phi^j~,
\eer{123}
where $E=g+B$.
The one-loop counterterm is proportional to
\ber
\Delta=\int d^2\xi ~d\theta ^+ d\theta^-  \left( (R^{(+)}_{ij} + \partial_{[i} \alpha_{j]})D_+\Phi^i D_-\Phi^j
- U^i \frac{\delta S} {\delta \Phi^i}
\right)~.
\eer{123a}
Here $R^{(+)}_{ij}$
is the Ricci tensor with torsion and the term involving $\alpha$ is a total derivative.
The term proportional to $U$ vanishes when the classical field equation ${\delta S}/ {\delta \Phi^i}=0$ is imposed, and off-shell can be absorbed into a field redefinition\footnote{There are more general field redefinitions which may involve 
 the derivatives $D_\pm$ and dimensionful parameters. These field redefinitions are not relevant at the given order. 
A similar comment is applicable to $(2,1)$ and $(2,2)$ superspace. } of $\Phi$: $\Phi \rightarrow \Phi + U(\Phi)$.
Integrating by parts (and shifting $\alpha$), this can be rewritten as
\ber
\Delta=\int d^2\xi ~d\theta ^+ d\theta^-  \left( R^{(+)}_{ij} + 2\nabla _{(i} U_{j)}+ H_{ijk} U^k+\partial_{[i} \alpha_{j]}\right)
 D_+\Phi^i D_-\Phi^j~,
\eer{1234}
or as
\ber
\Delta=\int d^2\xi ~d\theta ^+ d\theta^-  \left( R^{(+)}_{ij} + 
{\cal L}_U E_{ij}
 +\partial_{[i} \alpha_{j]}\right)
 D_+\Phi^i D_-\Phi^j
\eer{123b}
after a further shift of $\alpha$, where ${\cal L}_U$ is the Lie derivative with respect to $U$.
For a given geometry, one-loop finiteness requires that there is a choice of vector $U$ and 1-form
$\alpha$ such that \cite{Hull:1985rc}
\ber
R^{(+)}_{ij} + 2\nabla _{(i} U_{j)}+ H_{ijk} U^k+\partial_{[i} \alpha_{j]}=0~.
\eer{vanish}
For geometries without torsion, this gives the condition for finiteness
\ber
R_{ij} + 2\nabla _{(i} U_{j)}=0~.
\eer{sdfjklas}
For a K\"ahler manifold, this becomes
\ber
R_{\alpha\bar \beta}+2\nabla_{(\alpha}U_{\bar \beta)}=0~,~~
\nabla_{(\alpha}U_{\beta)}=\nabla_{(\bar \alpha}U_{\bar \beta)}=0~.
\eer{G33}

\subsection{The $(2,2)$ model without Torsion}

Before turning to the $(2,1)$ and $(2,2)$ supersymmetric cases with torsion, it will be useful to 
first review the case of $(2,2)$ sigma models without torsion, with K\"ahler target space.
The general one-loop counterterm in this case is
\ber
\Delta_{(2,2)}= \int d^2\xi~d^4\theta\left[\half \ln(\det(K_{\alpha\bar \alpha}))+
Z^\alpha(\boldsymbol{\phi})K_\alpha+\bar Z^{\bar \alpha}(\bar{\boldsymbol{\phi}})K_{\bar \alpha}\right]~,
\eer{F1}
where $Z^\alpha(\boldsymbol{\phi})$ is an arbitrary {\it holomorphic} vector field and corresponds 
to the field redefinitions
\ber
 \boldsymbol{\phi}^\alpha~\longrightarrow~\boldsymbol{\phi}^\alpha  +  Z^\alpha(\boldsymbol{\phi})~.
\eer{redchirlala}
The two last terms in (\ref{F1}) are proportional to the equations of motion.
The requirement for finiteness thus becomes
\ber
\half \ln(\det(K_{\alpha\bar \alpha}))+
Z^\alpha(\boldsymbol{\phi})K_\alpha+\bar Z^{\bar \alpha}(\bar{\boldsymbol{\phi}})K_{\bar \alpha}
=f(\boldsymbol{\phi})+\bar f(\bar{\boldsymbol{\phi}})~.
\eer{F2}

This condition must be compatible with 
the condition \re{G33} obtained earlier from the $(1,1)$ analysis.
Integrating over two of the fermionic coordinates to obtain a $(1,1)$ superspace form of this counterterm must then give a Ricci tensor term plus terms involving vector fields, so this implies that there must be an identity involving a relation between the Ricci tensor and an expression with two derivatives acting on
$\ln(\det(K_{\alpha\bar \alpha}))$.
There is indeed such an expression, the well-known identity for K\"ahler manifolds:
\ber
R_{\alpha\bar \alpha}=\partial_\alpha\bar\partial_{\bar \alpha}\ln(\det(K_{\beta\bar \beta}))~.
\eer{F4}
However, if this had not been known, the superspace arguments would have led us to discover it. This kind of argument  will  lead to interesting identities in the cases with torsion.
Then acting on  \re{F2}  with $\partial_\alpha\bar\partial_{\bar \alpha}$ 
and using the
 expression \re{F4} 
yields the condition for finiteness
\ber
R_{\alpha\bar \alpha}+2\nabla_{(\alpha}Z_{\bar \alpha)}=0~.
\eer{F3}
This is of the same form as the condition \re{G33} found above with $U=Z$, but has the extra restriction that the vector $Z$ is required to be holomorphic ($\bar\partial_{\bar \alpha}Z^\alpha=0$).
If equation \re{G33} is satisfied with a nonholomorphic $U$, the theory would be one-loop finite when regarded as a $(1,1)$ sigma model but the nonlinear wave function renormalisation required for off-shell finiteness does not respect
the full $(2,2)$ supersymmetry. We know of no examples where this happens.

\subsection{The $(2,1)$ Sigma Model}

Next we turn to the $(2,1)$ sigma-model \re{2}. The one-loop counterterm was given in \cite{Hull:1985zy}.
The one-loop renormalization of $\lambda_i$ is proportional to $\Gamma^{(+)}_j$ where
$\Gamma^{(+)}$ is the $U(1) $ part of the connection with torsion:
\ber
\Gamma^{(+)}_i = J^{~j}_{+k}\Gamma^{(+)k}_{ij} \, .
\eer{asd}
It can be written \cite{Hull:1985zy} in terms of a one form $v^{(+)}$
(known as the Lee form for $\omega^{(+)}$ in Hermitian geometry)
defined by 
\ber
v^{(+)}_i   
=J_{+i}^{~j}\nabla_k J^{~k}_{+j} 
= -\frac{1}{2}J_{+i}^{~j}H_{jk}^l J^{~k}_{+l} ~.
\eer{dfgdf}
(which vanishes if the torsion vanishes) and the determinant of the metric.
In complex coordinates adapted to $J_{+}$,
\ber
v^{(+)}_\alpha= - g^{\beta\bar \gamma} H_{\alpha\beta\bar \gamma}=  g^{\beta\bar \gamma}
 (g_{\alpha\bar \gamma, \beta}-g_{\beta\bar \gamma, \alpha})
\eer{dfgert}
and
\ber
\Gamma^{(+)}_\alpha = i\left(2v^{(+)}_\alpha + \partial_\alpha\ln{\det g_{\beta \bar \gamma}}\right)~.
\eer{sflskdjf}
The curvature of the $U(1)$ part of the connection is
\ber
C_{ij}^{(+)}=\partial _{i}\Gamma^{(+)}_{j}-\partial _{j}\Gamma^{(+)}_{i}~.
\eer{fdgdrt}

Allowing for terms that vanish on-shell and total derivatives, the general form of the one-loop counterterm is
\ber
S=-\frac i2 \int d^2\xi~d\theta ^+d\bar\theta^+d\theta^-(\Gamma^{(+)}_i +
 {\cal L} _{V^{(+)}} \lambda^{(+)}_i+\partial _i \rho^{(+)} ) D_-\varphi^i ~.
\eer{2ssa} 
Here  ${\cal L} _{V^{(+)}}$ denotes the Lie derivative with respect to a  vector field
$(V^{(+)})^i= ((V^{(+)})^\alpha, (\bar V^{(+)})^{\bar \alpha})$, with $(V^{(+)})^\alpha(\boldsymbol{\varphi})$
a holomorphic vector field.
The term involving $\rho^{(+)} _i$ is a total derivative, included for generality.
The term involving $V^{(+)}$ again vanishes on-shell (up to a surface term).
The condition for one-loop finiteness is then
\ber
\Gamma^{(+)}_i + {\cal L} _{V^{(+)}} \lambda^{(+)}_i +\partial _i \rho^{(+)} 
= f_i(\boldsymbol{\varphi})+\bar f_i(\bar{\boldsymbol{\varphi}})~.
\eer{sdfs}

The results from $(2,1)$ superspace must be compatible with those from $(1,1)$ superspace.
In particular, the $(2,1)$ counterterm \re{2ssa}  with $V^{(+)}=0$ 
and $\rho^{(+)} =0$ gives a $(1,1)$ superspace counterterm
involving derivatives of $\Gamma^{(+)}_i $.
This must agree with the counterterm
\re{123} for some choice of $U,\alpha$, and for this to be the case, there must be 
some identities relating $R^{(+)}_{ij}$ to derivatives of $\Gamma^{(+)}_i $. 
This is is indeed the case, and leads to the remarkable identities 
\cite{Hull:1985zy},\cite{Hull:1986iu},\cite{Hull:1986kz}
\ber
R_{\alpha\beta}^{(+)} &=& \nabla_\alpha^{(-)} v^{(+)}_\beta~,
\\
\label{eqcomep33} R_{\alpha\bar \beta}^{(+)} &=& \nabla_\alpha^{(-)}v^{(+)}_{\bar \beta}
-\frac{i}{2} C^{(+)}_{\alpha\bar \beta}- (\partial_\alpha v^{(+)}_{\bar \beta}- \partial _{\bar \beta} v^{(+)}_\alpha)
~,
\eer{eqcomep22}
which may be written covariantly as
\ber
R^{(+)}_{ik} = \nabla_i^{(-)} v_{k}^{(+)} - \frac{1}{2}\left(J_+C^{(+)}\right)_{ik} - \left( dv^{(+)} \right)_{ik}
\eer{compcovp}
which imply consistency between \re{eqcomep22} and \re{vanish} with
\ber
U_i=-\frac12 v^{(+)}_i, \qquad \alpha_i =  v^{(+)}_i~.
\eer{erewter}

\subsection{The $(1,2)$ Sigma model}
For completeness we give the corresponding formulas for the $(1,2)$ sigma model.
The $U(1)$ part of the connection is
\ber
\Gamma^{(-)}_i = J^{j}_{-k}\Gamma^{(-)k}_{ij},
\eer{}
which can be written in terms of the Lee-form $v^{(-)}$
\ber
v^{(-)}_i = J_{-i}^j\nabla_k J_{-j}^k = \frac12 J_{-i}^j H^l_{jk}J_{-l}^k,
\eer{}
and the determinant of the metric. In complex coordinates adapted to $J_-$
we have
\ber
v^{(-)}_\alpha = g^{\beta\bar \gamma} H_{\alpha\beta\bar \gamma} = 
-g^{\beta\bar \gamma}(g_{\alpha\bar \gamma,\beta}-g_{\beta\bar \gamma,\alpha}),
\eer{}
and
\ber
\Gamma^{(-)}_{\alpha} = i(2v^{(-)}_\alpha+\partial_\alpha \ln\det g_{\beta\bar \gamma})
\eer{}
The curvature of the $U(1)$ part of the connection is
\ber
C^{(-)}_{ij} = \partial_i\Gamma^{(-)}_j -\partial_j \Gamma^{(-)}_i
\eer{}
The one-loop counterterm is given by
\ber
S = -\frac i2 \int d^2\xi d\theta^- d\bar\theta^- d\theta^+(\Gamma^{(-)} + 
{\cal L}_{V^{(-)}} \lambda^{(-)} + \partial_i\rho^{(-)})D_+\varphi^i~.
\eer{2ssam}
The relation between the $U(1)$ curvature and the Ricci tensor is
\ber
R^{(-)}_{ik} = \nabla^{(+)}_i v^{(-)}_k -\frac12\left(J_-C^{(-)}\right)_{ik}
-\left(dv^{(-)}\right)_{ik}
\eer{compcovm}

%%%%%%%%
\subsection{The $(2,2)$ model with Torsion}

We turn now to the $(2,2)$ case. 
The one-loop counterterm is proportional to
\ber 
\int d^2\xi~d^4\theta\left[  K^1+{\cal L}_W K\right]~,
\eer{FF6}
where 
$K^1$ is the one-loop counterterm calculated in \cite{Buscher:1985kb}, given by
\ber 
K^1=\ln\left( \frac {A}{B}\right)~,
\eer{k1is}
where $A$ and $B$ are given in (\ref{F7}) below.

The results from $(2,2)$ superspace must be compatible with the results in $(2,1)$ and
$(1,2)$ as well as the result in $(1,1)$ superspace. In particular, the $(2,2)$ counterterm
(\ref{FF6}) with $W=0$ reduces to the counterterms (\ref{2ssa}) and (\ref{2ssam}) with in general
nonzero $V^{(\pm)}$ and $\rho^{(\pm)}$.

We can investigate in which cases we get nonzero $V^{(\pm)}$ by reducing the two counterterms
further to $(1,1)$ superspace. If we had $V^{(\pm)}=0$ we would get the $(1,1)$
counterterms
\ber
-\frac12\int d^2\xi d^2\theta D_+\phi^i \left( J_+C^{(+)}\right)_{ik} D_-\phi^k
\eer{}
and
\ber
\frac12\int d^2\xi d^2\theta D_+\phi^i \left(C^{(-)}J_-\right)_{ik} D_-\phi^k
\eer{}
Since they both come from the same $(2,2)$ lagrangian $K^1$ we know that they should
differ by a closed two-form
\ber
J_+C^{(+)}+C^{(-)}J_- = d(K^1_T d\chi-K^1_C d\phi)
\eer{}
But from (\ref{compcovp}) and (\ref{compcovm}) and the fact that $R^{(+)}_{ik} = R^{(-)}_{ki}$
we know that
\ber
J_+C^{(+)}+C^{(-)}J_- = dX+{\cal L}_Y g_{ik} + H^l_{ik}Y_l
\eer{}
where
\ber
X &=& v^{(+)} + v^{(-)}\\
Y &=& v^{(+)} - v^{(-)}
\eer{}
which proves that $V^{(\pm)} \neq 0$ when $Y \neq 0$. As we will see in the next section, in
the case of commuting complex structures, $v^{(+)} = v^{(-)}$.

\subsubsection{$[J_{+},J_{-}]=0$}

When $[J_{+},J_{-}]=0$, there are only chiral and twisted chiral superfields 
and the  $(2,2)$ superspace Lagrangian is given by a potential
\ber
K=K(\boldsymbol{\phi}, \bar{\boldsymbol{\phi}}, \boldsymbol{\chi}, \bar{\boldsymbol{\chi}})~.
\eer{F555}
On converting to $(2,1)$ superspace, this gives a potential
\ber
\lambda^{(+)} = -J_{-} dK =  d^c_- K~.
\eer{jsdfl}
The $(2,2)$ one-loop counterterm (\ref{FF6})
where
\ber 
A =\det ( -K_{t\bar{t}}), \qquad B= \det (K_{c\bar{c}})
\eer{POP}
and the vector $W$ is holomorphic with respect to both $J_{\pm}$:
\ber
W=(W^c (\boldsymbol{\phi}), \bar W^{\bar c}(\bar{\boldsymbol{\phi}}), W^t  (\boldsymbol{\chi}), \bar W^{\bar t}  (\bar{\boldsymbol{\chi}}))~.
\eer{wiss}
gives a $(2,1)$ one-loop counterterm with
\ber
\lambda^{+1}=- J_{-} dK^1~.
\eer{jsdfl1}

Note that 
\ber
\sqrt{\det (g_{ij})} = AB
\eer{dett}
and $v^{(+)}$ is given by 
\ber
v^{(+)}_c= - \partial _c \ln A~, \qquad v^{(+)}_t=- \partial _t \ln B~,
\eer{dfgdffgh}
so that using \re{sflskdjf}, $\Gamma^{(+)}_a $ is given by 
\ber
\Gamma^{(+)}_c =i   \left(   - 2\partial _c \ln A + \partial_c \ln(AB) \right)=
-i     \partial_c \ln(A/B)  
\eer{sflskdjfs}
and 
\ber
\Gamma^{(+)}_t = i   \left(   - 2\partial _t \ln B + \partial_t \ln(AB) \right)=
i \partial_t \ln(A/B)  ~.
\eer{sflskdjfsfh}
As a result 
\ber
\Gamma^{(+)}=\lambda ^{+1}= -J_{-} dK^1
\eer{jsdflsf}
and the $(2,2)$ counterterm \re{FF6} with $W=0$ gives the $(2,1)$ counterterm 
\re{2ssa}  with $V^{(+)}=0, \rho^{(+)}=0$.

Similarly, going to $(1,2)$ superspace we have
\ber
\lambda^{(-)} = J_+dK = -d_+^cK
\eer{}
and the one loop counterterm becomes
\ber
\lambda^{-1} = J_+dK^1~.
\eer{}
In this case we have
\ber
v^{(-)} &=& v^{(+)}\\
\Gamma^{(-)}_c &=& -i\partial_c\ln(A/B)\\
\Gamma^{(-)}_t &=& -i\partial_t\ln(A/B)
\eer{}
As a result
\ber
\Gamma^{(-)} = -\lambda^{-1} = -J_+dK^1
\eer{}

Note that if the generalized Monge-Amp\`ere equation \cite{Hull:2010sn}
\ber
A=B
\eer{maa}
is satisfied, then
$K^1=0$ and
\ber
v^{(\pm)}_i= -2\partial _i \Phi~,
\eer{vissd}
where
\ber
\Phi =-2 \ln A
\eer{htfg}
 is the dilaton (not to be confused with a $(1,1)$ superfield).

\subsubsection{General $[J_{+},J_{-}]\ne0$}

In the general case, the $(2,2)$ superspace Lagrangian is given by the potential
\ber
K=K(\boldsymbol{\phi}, \bar{\boldsymbol{\phi}}, \boldsymbol{\chi}, \bar{\boldsymbol{\chi}}, 
\boldsymbol{X}_{l}, \bar{\boldsymbol{X}}_{\bar l},  \boldsymbol{X}_r,\bar{\boldsymbol{X}}_{\bar r})~.
\eer{F5}
The one-loop counterterm is then proportional to
\ber\nn
&&\int d^2\xi~ d^4\theta\left[  K^1+W^c(\boldsymbol{\phi})K_c+\bar W^{t}(\boldsymbol{\chi})K_{t}+W^{l}(\boldsymbol{\phi}, \boldsymbol{\chi}, \boldsymbol{X}_{l})K_{l}+
W^{r}(\boldsymbol{\phi}, \bar{\boldsymbol{\chi}}, \boldsymbol{X}_{r})K_{r}+c.c.\right]~,\\[1mm]
&&
\eer{F6}
where 
$K^1$ is the one-loop counterterm calculated in \cite{Grisaru:1997pg}. It was shown in \cite{Hull:2010sn} that $K^1$ can be rewritten as
\ber 
K^1=\ln\left( \frac {A}{B}\right)~,
\eer{k1is1}
where
\ber\nn
& A= 
  \det \left ( \begin{array}{ccc}
  - K_{l\bar{l}} & - K_{lr} & - K_{l\bar{t}} \\
  - K_{\bar{r}\bar{l}} & - K_{\bar{r}r} & - K_{\bar{r}\bar{t}} \\
  - K_{t\bar{l}} & -K_{tr} & -K_{t\bar{t}} \end{array} \right ) ~,\\[2mm]
& B= \det \left (\begin{array}{ccc}
   K_{l\bar{r}} & K_{l\bar{l}} & K_{l\bar{c}} \\
   K_{r\bar{r}} & K_{r\bar{l}} & K_{r\bar{c}} \\
    K_{c\bar{r}} & K_{c\bar{l}} & K_{c\bar{c}}
    \end{array} \right )~.
 \eer{F7}
 The condition for one-loop finiteness is then
 \ber\nn
 &&K^1+W^c(\boldsymbol{\phi})K_c+\bar W^{t}(\boldsymbol{\chi})K_{t}+W^{l}(\boldsymbol{\phi}, \boldsymbol{\chi}, \boldsymbol{X}_{l})K_{l}+
W^{r}(\boldsymbol{\phi}, \bar{\boldsymbol{\chi}}, \boldsymbol{X}_{r})K_{r}+c.c.\qquad\qquad\qquad\\[1mm]
&&\qquad\qquad=f^+(\boldsymbol{\phi},\boldsymbol{\chi},\boldsymbol{X}_{l})+\bar f^+(\bar{\boldsymbol{\phi}},\bar{\boldsymbol{\chi}},
\bar{\boldsymbol{X}}_{\bar l})+
f^-(\boldsymbol{\phi},\bar{\boldsymbol{\chi}}, \boldsymbol{X}_{r})+\bar f^-(\bar{\boldsymbol{\phi}}, \boldsymbol{\chi},\bar{\boldsymbol{X}}_{\bar r})~.
\eer{F8}
Note that the determinant of the metric in these coordinates is \cite{Hull:2010sn}
 \ber
\sqrt{\det g_{\mu\nu}} = \frac{(-1)^{d_sd_c}}{\det K_{LR}}
 A B \eer{detg}
or in coordinates adapted to either of the complex structures is
\ber
\det g_{a\bar b} = \frac{1}{(\det K_{LR})^2}
 AB~.
 \eer{detcc}
The  counterterm \re{F6}
for any given choice of the vector field $W$ (including $W=0$) must give a counterterm in $(1,1)$ superspace of the form \re{123} for some definite $U, \alpha$ (which will depend on the choice of $W$).
 
\subsection{Renormalization}

We can also compare with the $(2,1)$ superspace counterterm \re{2ssa}.
Note that there is a subtlety here:
Quantization in $(2,1)$ superspace preserves $J_{+}$ and the chiral constraints, so 
that the ambiguity involves a holomorphic vector field $V^{(+)}$.
In $(2,2)$ superspace, the complex structures are implicitly defined in terms of the
holomorphic coordinates, for $J_+$ they are ${\boldsymbol \phi};{\boldsymbol \chi};
{\boldsymbol X}_l; {\boldsymbol Y}_l = \partial K/\partial {\boldsymbol X}_l$.
Quantization in $(2,2)$ superspace  preserves the structure leading to the superfield constraints 
(the semichirality constraints on $X_{L}$ and $X_R$) but does {\em not} preserve $J_{+}$.
This means that for models with semichiral superfields $X$ in $(2,2)$ superspace,
the counterterm in $(2,1)$ superspace arising from the $(2,2)$ counterterm differs from the 
the $(2,1)$ counterterm \re{2ssa} by a choice of $V^{(+)}$ that is incompatible with holomorphy,
and the two results can be reconciled only by descending all the way to $(1,1)$ 
superspace.
\footnote{This may seem surprising, but there is an analog in the usual $(2,2)$ 
chiral superfield description of hyperk\"ahler manifolds: the $(2,2)$ Lagrangian 
is the K\"ahler potential with respect to a given
complex structure; changing the complex structure leads to a different K\"ahler potential that
cannot be related to the original by any holomorphic coordinate redefinition, but descending to 
$(1,1)$ superspace, one may find a real coordinate redefinition that shows the equivalence of
the apparently different $(2,2)$ models.}

The one-loop quantum theory in $(2,2)$ superspace is given in terms of a potential
\ber 
\hat K = K+ \hbar t K^1 + O(\hbar^2)~,
 \eer{quanst}
where $K$ is the classical potential, $\hbar$ is the loop-counting parameter and $t$ is a scale-dependent term.
(Here $t $ is proportional to $\log (\mu^2/m^2)$ where $\mu$ is the ultraviolet regularisation mass scale and $m$ is the infrared regularisation mass scale.)
To find the $(2,1)$ superspace form, we use the results from sections 3 and 4 and use $\hat K$ instead of $K$ in equations \re{3333},\re{7b},\re{7c},\re{ni},\re{r7}\re{recov}.
For the one-loop corrections, we seek the terms linear in $\hbar$. The change $K^1$ to the potential
does not change
the definition of the fields ${X}_{L,R},\phi,\chi$; however, it does
change the $J_{+},J_{-}$ holomorphic fields $Y_L,Y_R$, respectively, and hence it renormalizes the 
complex structures $J_{\pm}$ (when the model has semichiral fields, {\it i.e.}, when $[J_{+},J_{-}]\ne0$).
The reason why the holomorphic fields $Y$ change is that they are defined through derivatives of $K$ which
changes. For instance
\ber
\hat Y_L = \frac{\partial \hat K}{\partial X_L} = Y_L + \hbar t Y^{(1)}_L + O(\hbar ^2)~,
\eer{dfgd}
where
\ber 
Y^{(1)}_L = K^1_L~.
\eer{fdsgh}

\section{Conclusions}
We have investigated the role of $(2,1)$ superspace in generalized K\"ahler geometry.
In particular we have discussed the treatment of the one-loop quantum corrections
in $(2,1)$ superspace. The main advantage of $(2,1)$ superspace is that
all fields are chiral in contrast to the multitude of superfields necessary to describe
the most general sigma-model in $(2,2)$ superspace.
It is interesting to notice that the one-loop counterterm
is proportional to the $U(1)$ connection of the target space geometry.
We have shown how the counterterms reduces when one integrates out part of the
superspace coordinates showing the necessity of nontrivial wave function
renormalization to reconcile the results.
An open question is how to express the $U(1)$ connection in terms of the
generalized K\"ahler potential.

\bigskip\bigskip
\noindent{\bf\Large Acknowledgement}:
\bigskip\bigskip

\noindent

We thank Nigel Hitchin for discussions. 
We are grateful to  
Simons Center where part of this work was carried out,
for providing a stimulating atmosphere.
 We are also grateful to the program ``Geometry of strings and fields" at Nordita 
  where part of this work was carried out,
for providing a stimulating atmosphere.
The research of UL was supported by  VR grant 621-2009-4066.
The research of MR  was supported in part by NSF grant no.~PHY-06-53342.
The research of R.v.U. was
supported by the Grant 
agency of the Czech republic under the grant P201/12/G028..
The research of M.Z. was supported by VR-grants  621-2008-4273 and 621-2011-5079.


\begin{thebibliography}{6666}

\newcommand{\np}{{\em Nucl.\ Phys.\ }}
\newcommand{\pr}{{\em Phys.\ Rev.\ }}
\newcommand{\cmp}{{\em Commun.\ Math.\ Phys.\ }}
\newcommand{\pl}{{\em Phys.\ Lett.\ }}

%\cite{Hitchin:2004ut}
\bibitem{Hitchin:2004ut}
  N.~Hitchin,
  {\it Generalized Calabi-Yau manifolds,}
  Quart.\ J.\ Math.\ Oxford Ser.\  {\bf 54}, 281-308 (2003).
  [math/0209099 [math-dg]].
%%CITATION = MATH/0209099;%%

%\cite{Gualtieri:2003dx}
\bibitem{Gualtieri:2003dx}
  M.~Gualtieri,
  {\it Generalized complex geometry,}
  Oxford University DPhil thesis,
  [arXiv:math/0401221].
%%CITATION = MATH/0401221;%%

%\cite{Lindstrom:2005zr}
\bibitem{Lindstrom:2005zr}
  U.~Lindstr\"om, M.~Ro\v{c}ek, R.~von Unge and M.~Zabzine,
  {\it Generalized K\"ahler manifolds and off-shell supersymmetry,}
  Commun.\ Math.\ Phys.\  {\bf 269}, 833 (2007)
  [arXiv:hep-th/0512164].
%%CITATION = CMPHA,269,833;%%

%\cite{Hull:2008vw}
\bibitem{Hull:2008vw}
  C.~M.~Hull, U.~Lindstr\"om, M.~Ro\v{c}ek , R.~von Unge and M.~Zabzine,
  {\it Generalized K\"ahler geometry and gerbes,}
  JHEP {\bf 0910}, 062 (2009).
  [arXiv:0811.3615 [hep-th]].
%%CITATION = ARXIV:0811.3615;%%

%\cite{Hull:1985jv}
\bibitem{Hull:1985jv}
  C.~M.~Hull and E.~Witten,
  {\it Supersymmetric Sigma Models and the Heterotic String,}
  Phys.\ Lett.\ B {\bf 160} (1985) 398.
  %%CITATION = PHLTA,B160,398;%%

%\cite{Hull:1985zy}
\bibitem{Hull:1985zy}
  C.~M.~Hull,
  {\it Sigma Model Beta Functions And String Compactifications,}
  Nucl.\ Phys.\  B {\bf 267} (1986) 266.
%%CITATION = NUPHA,B267,266;%%

%\cite{Dine:1986by}  
\bibitem{Dine:1986by}
  M.~Dine and N.~Seiberg,
  {\it ``(2,0) Superspace,}
  Phys.\ Lett.\ B {\bf 180} (1986) 364.
  %%CITATION = PHLTA,B180,364;%%

%\cite{Howe:1987qv}
\bibitem{Howe:1987qv}
  P.~S.~Howe and G.~Papadopoulos,
  {\it Ultraviolet Behavior Of Two-dimensional Supersymmetric Nonlinear Sigma Models,}
  Nucl.\ Phys.\ B {\bf 289} (1987) 264.
  %%CITATION = NUPHA,B289,264;%%
  
%\cite{Howe:1988cj}
\bibitem{Howe:1988cj}
  P.~S.~Howe and G.~Papadopoulos,
  {\it Further Remarks On The Geometry Of Two-dimensional Nonlinear Sigma Models,}
  Class.\ Quant.\ Grav.\  {\bf 5} (1988) 1647.
  %%CITATION = CQGRD,5,1647;%%
  
%%\cite{AbouZeid:1997cw}
\bibitem{AbouZeid:1997cw}
  M.~Abou Zeid and C.~M.~Hull,
  {\it The gauged (2,1) heterotic sigma model,}
  Nucl.\ Phys.\  B {\bf 513}, 490 (1998)
  [arXiv:hep-th/9708047].
%%CITATION = NUPHA,B513,490;%%

%\cite{Hull:2010sn}
\bibitem{Hull:2010sn}
  C.~M.~Hull, U.~Lindstr\"om, M.~Ro\v{c}ek, R.~von Unge, M.~Zabzine,
  {\it Generalized Calabi-Yau metric and Generalized Monge-Ampere equation,}
  JHEP {\bf 1008}, 060 (2010).
  [arXiv:1005.5658 [hep-th]].
%%CITATION = ARXIV:1005.5658;%%

%\cite{Gates:1984nk}
\bibitem{Gates:1984nk}
  S.~J.~Gates, Jr., C.~M.~Hull, M.~Ro\v{c}ek,
  {\it Twisted Multiplets and New Supersymmetric Nonlinear Sigma Models,}
  Nucl.\ Phys.\  {\bf B248}, 157 (1984).
%%CITATION = NUPHA,B248,157;%%

%\cite{Buscher:1987uw}
\bibitem{Buscher:1987uw} 
 T.~Buscher, U.~Lindstrom and M.~Rocek,
 {\it New supersymmetric sigma models with Wess Zumino terms,}
 Phys.\ Lett.\ B {\bf 202}, 94 (1988).
 %%CITATION = PHLTA,B202,94;%%

%\cite{Hull:1991uw}
\bibitem{Hull:1991uw}
 C.~M.~Hull, G.~Papadopoulos and B.~J.~Spence,
{\it Gauge symmetries for (p,q) supersymmetric sigma models,}
 Nucl.\ Phys.\ B {\bf 363} (1991) 593.
%%CITATION = NUPHA,B363,593;%%

%\cite{Lindstrom:2007vc}
\bibitem{Lindstrom:2007vc}
  U.~Lindstr\"om, M.~Ro\v{c}ek, I.~Ryb, R.~von Unge, M.~Zabzine,
  {\it New N = (2,2) vector multiplets,}
  JHEP {\bf 0708}, 008 (2007).
  [arXiv:0705.3201 [hep-th]].
%%CITATION = ARXIV:0705.3201;%%

%\cite{Lindstrom:2008hx}
\bibitem{Lindstrom:2008hx}
  U.~Lindstr\"om, M.~Ro\v{c}ek, I.~Ryb, R.~von Unge, M.~Zabzine,
  {\it Nonabelian Generalized Gauge Multiplets,}
  JHEP {\bf 0902}, 020 (2009).
  [arXiv:0808.1535 [hep-th]].
%%CITATION = ARXIV:0808.1535;%%

%\cite{Hull:1985rc}
\bibitem{Hull:1985rc}
  C.~M.~Hull and P.~K.~Townsend,
  {\it Finiteness And Conformal Invariance In Nonlinear Sigma Models,}
  Nucl.\ Phys.\  B {\bf 274} (1986) 349.
%%CITATION = NUPHA,B274,349;%%


%\cite{Hull:1986iu}
\bibitem{Hull:1986iu}
  C.~M.~Hull,
 {\it Superstring Compactifications With Torsion And Space-Time
Supersymmetry,} in the Proceedings of the First Torino Meeting on Superunification and
Extra Dimensions, edited by R.~D'Auria and P.~Fre, (World Scientific,
Singapore, 1986).
%%CITATION = PRINT-86-0251-CAMBRIDGE-;%%

%\cite{Hull:1986kz}
\bibitem{Hull:1986kz}
  C.~M.~Hull,
  {\it Compactifications of the Heterotic Superstring,}
  Phys.\ Lett.\  B {\bf 178} (1986) 357.
%%CITATION = PHLTA,B178,357;%%

%\cite{Buscher:1985kb}
\bibitem{Buscher:1985kb}
  T.~H.~Buscher,
  {\it Quantum Corrections And Extended Supersymmetry In New Sigma Models,}
  Phys.\ Lett.\  B {\bf 159}, 127 (1985).
%%CITATION = PHLTA,B159,127;%%

%\cite{Grisaru:1997pg}
\bibitem{Grisaru:1997pg}
  M.~T.~Grisaru, M.~Massar, A.~Sevrin and J.~Troost,
  {\it The quantum geometry of N = (2,2) nonlinear sigma-models,}
  Phys.\ Lett.\  B {\bf 412} (1997) 53
  [arXiv:hep-th/9706218].
%%CITATION = PHLTA,B412,53;%%

\end{thebibliography}
\end{document}